\documentclass[aps,showpacs,twocolumn,prb,amsmath,amssymb,superscriptaddress,intlimits,longbibliography,showkeys]{revtex4-1}

\usepackage{graphicx}
\usepackage{subfigure}
\usepackage{xcolor}
\usepackage{epstopdf}
\usepackage{amssymb}
\usepackage{amsmath}
\usepackage{amsfonts}
\usepackage{dcolumn}
\usepackage{bm}
\usepackage{soul}
\usepackage{array}
\usepackage{physics}

\newcommand{\vQ}{{\bf Q}}
\newcommand{\vq}{{\bf q}}

\newcolumntype{C}[1]{>{\centering\let\newline\\\arraybackslash\hspace{0pt}}m{#1}}

\graphicspath{{figures/}}

\begin{document}

\title{Exciton valley depolarization in monolayer MoS$_2$: non-Markovian quantum dynamics, intervalley scattering, and the breakdown of the Dyakonov-Perel mechanism}

\author{Yang-hao Chan}
\email{yanghao@gate.sinica.edu.tw}
\affiliation{Institute of Atomic and Molecular Sciences, Academia Sinica, Taipei 10617, Taiwan}

\author{Jonah B. Haber}
\affiliation{Department of Materials Science and Engineering, Stanford University, Stanford, CA 94305, USA}

\author{Mit H. Naik}
\affiliation{Department of Physics, University of Texas at Austin, Austin, TX, 78712, USA}

\author{Felipe H. da Jornada}
\email{jornada@stanford.edu}
\affiliation{Department of Materials Science and Engineering, Stanford University, Stanford, CA 94305, USA}

\author{Diana Y. Qiu}
\email{diana.qiu@yale.edu}
\affiliation{Department of Materials Science, Yale University, New Haven, CT 06520}

\begin{abstract}

In monolayer transition metal dichalcogenides, exciton valley relaxation is typically attributed to the Dyakonov-Perel (DP) mechanism, where frequent intravalley scattering with phonons or defects suppresses the intervalley exchange-induced precession and the role of intervalley scattering is considered secondary.
Employing a first-principles nonequilibrium exciton Green's function (NEGF) approach with the generalized Kadanoff-Baym ansatz (GKBA), which treats exchange-driven precession and exciton-phonon scattering on an equal footing across the full Brillouin zone, we demonstrate that even in the small momentum regime most favorable to DP physics, realistic exciton-phonon scattering is too weak to induce the motional narrowing that defines the DP regime. 
Upon accounting for excitons across the entire Brillouin zone, large momentum intervalley scattering becomes the dominant pathway for valley relaxation, shortening the depolarization by a factor of 3-4 relative to intravalley-only models.
We find that valley depolarization and decoherence time are approximately 50 fs at 300 K and lengthen to 130 fs at 10 K.
By comparing our results with a Lindblad-type collision framework, we explicitly demonstrate the role of non-Markovian effects and reveal that the Markovian Lindblad framework is highly basis dependent. In the valley-pseudospin basis underlying prior analyses, the Lindblad approach artificially amplifies scattering-induced equilibration, biasing the dynamics toward a DP interpretation.
Our study provides a comprehensive picture of valley relaxation, and establishes the exciton density matrix approach derived from NEGF+GKBA as a powerful tool for investigating ultrafast exciton dynamics.

\end{abstract}

\date{\today}
\maketitle

\section{Introduction}

Valley physics in low-dimensional transition metal dichalcogenides (TMDs) is of particular interest owing to their unique valley-spin locking~\cite{Cao2012,Xiao2012} in K and K' valleys connected by time-reversal symmetry and their experimental accessibility facilitated by large exciton binding energies~\cite{Qiu2013,Ugeda2014,Chernikov2014,Qiu2016,Wang2018}.
The capability of selectively addressing a single valley with circularly polarized light~\cite{Zeng2012,Mak2012,Sallen2012} opens prospects for valleytronic and quantum science applications~\cite{Xu2014,Schaibley2016,Ye2017,Mueller2018,Zhao_2021}.

A critical metric for evaluating a material's viability in valleytronics is the valley relaxation time. 
Interactions known for driving valley relaxation in TMDs include phonon-assisted intervalley electron scattering~\cite{Molina-Sanchez2017,Selig2019,Xu2021,Chen2022,Dogadov2026}, Coulomb-interaction based intervalley electron-hole exchange and Dexter-like interaction~\cite{Yu2014,Zhu2014,DalConte2015,Hao2016}, and many-body correlations~\cite{Mai2014,Mai2014_b,DalConte2015,Schmidt2016,Berghaeuser2018}.
By drawing an analogy to spin relaxation physics, with the $K$ and $K'$ valleys in TMDs treated as pseudospins, the Dyakonov-Perel mechanism (DP)~\cite{DP1971} (or Maialle-Silva-Sham (MSS) mechanism in the exciton context~\cite{Maialle1993}) has been widely adopted to describe ultrafast valley relaxation. In this picture, the long-range exchange interaction acts as an effective momentum-dependent magnetic field that drives valley pseudospin precession.
A hallmark of the DP regime is that the valley relaxation rate scales inversely with the momentum scattering rate~\cite{DP1971,Maialle1993,Park2022}.
This counterintuitive behavior originates from the motional narrowing of valley precession: frequent intravalley scattering from phonons or defects continuously resets the precession axis, thus averaging out the precession and slowing depolarization.

Despite its widespread use, the microscopic validity of the DP picture in realistic monolayer TMDs remains unsettled.
Ultrafast optical spectroscopies, such as transient Kerr and Faraday rotation spectroscopy~\cite{Zhu2014,DalConte2015}, time-resolved photoluminescence~\cite{Korn2011,Glazov2014,Lagarde2014,Wang2015}, helicity-resolved pump-probe experiments~\cite{Wang2013,Mai2014_b,DalConte2015,Dogadov2026}, and four-wave mixing~\cite{Hao2016,Jakubczyk2019,Purz2022} have been widely applied to study valley relaxation dynamics.
Although the DP mechanism has been invoked to explain these experiments, these previous studies rely on simplified models that typically neglect large-momentum intervalley scattering processes~\cite{Glazov2014,Zhu2014,Hao2016}.
In contrast, recent broadband femtosecond transient circular dichroism (CD) spectroscopy~\cite{Dogadov2026} combined with model simulations identifies intervalley electron-phonon scattering as the primary relaxation channel during the sub-picosecond regime.
These results point to an important physical question: is valley relaxation in monolayer TMDs governed primarily by coherent exchange-driven precession near the light cone, or by exciton-phonon scattering across the full exciton Brillouin zone?

Answering this question requires a theory that goes beyond exciton population dynamics: it must retain valley coherences in the exciton density matrix while treating exchange-driven precession and exciton-phonon scattering on the same footing. Existing first-principles approaches have provided important insights, but each leaves out part of the relevant physics~\cite{Jian2021,Molina-Sanchez2017,Chen2022}.
Specifically, the semiclassical Boltzmann framework neglects off-diagonal exciton coherences, which are fundamental to valley precession dynamics~\cite{DP1971,Maialle1993,Mower2011,Hao2016} and thus cannot fully capture the phase-dependent evolution of the valley pseudospin. Conversely, previous electron density matrix approaches naturally retain off-diagonal coherences but omit, by construction, the long-range intervalley exchange interaction. This interaction vanishes at the zone center and is only finite for excitons with non-zero center-of-mass momentum (COM)~\cite{Qiu2015}, making an exciton-basis treatment with full COM momentum resolution essential.

In this paper, we apply a first-principles nonequilibrium exciton Green's function (NEGF) approach~\cite{NEQ2013} with the generalized Kadanoff-Baym ansatz (GKBA)~\cite{Lipavsky1986} to study exciton valley dynamics in monolayer MoS$_2$.
By including exciton-phonon scattering and Coulomb-based exchange interaction on an equal footing, we resolve their distinct contributions to valley relaxation dynamics across the entire Brillouin zone (BZ). 
We first use simulations restricted to small-momentum excitons to establish the dynamical signatures of the DP mechanism. These calculations reveal that, near the zone center, the intervalley exchange interaction drives the relaxation, but physically realistic exciton-phonon coupling is insufficient to place the system in the DP regime. Notably, DP motional narrowing only appears when the exciton-phonon coupling strength is artificially enhanced, regardless of the temperature.
When excitons across the full BZ are included, large-momentum intervalley exciton-phonon scattering becomes the dominant valley relaxation pathway, shortening the depolarization time by a factor of 3-4 relative to zone-center-restricted models.

Our work also clarifies the theoretical requirements for simulating ultrafast exciton valley dynamics. We compare the GKBA dynamics with a Lindblad approach and show that the latter is basis dependent: formulated in the valley-pseudospin basis underlying prior analyses, it overestimates scattering-induced equilibration and produces a spurious motional-narrowing signature at realistic coupling strengths. In contrast, the GKBA formulation collision integral is covariant under basis transformations, ensuring representation-independent dynamics, while additionally incorporating non-Markovian scattering effects.
These non-Markovian effects manifest as rapid transient oscillations in the population of high-energy exciton states during the initial stage of dynamics, which may be accessible to attosecond spectroscopies~\cite{Lucchini2021,Schumacher2023}. 
At longer times, our calculated valley depolarization and decoherence times are approximately  130 fs at 10 K, in good agreement with experimental measurements~\cite{Dogadov2026, Hao2016}.

\section{Method}
\label{sec:method}

\subsection{Equation of motion}

We start by writing down the effective exciton Hamiltonian including exciton-phonon coupling~\cite{Antonius2022,Chen2020},
\begin{align}
H=&\sum_{n\mathbf{Q}}E_{n\mathbf{Q}}a_{n\mathbf{Q}}^{\dagger}a_{n\mathbf{Q}}+\sum_{\nu\mathbf{q}}\hbar\omega_{\nu\mathbf{q}}b_{\nu\mathbf{q}}^{\dagger}b_{\nu\mathbf{q}}\nonumber\\
+&\frac{1}{\sqrt{N_\vq}}\sum_{nm\nu,\mathbf{Q}\mathbf{q}}\mathcal{G}_{nm\nu}\left(\mathbf{Q},\mathbf{q}\right)a_{m\mathbf{Q}+\mathbf{q}}^{\dagger}a_{n\mathbf{Q}}\left(b_{\nu\mathbf{q}}+b_{\nu,-\mathbf{q}}^{\dagger}\right)\label{eq:Ham},
\end{align}
where $N_q$ is the number of $\vq$ points sampled in the BZ, $a^\dagger_{n\vQ}$ is a creation operator for an exciton with COM $\vQ$ and state $n$, $E_{n\vQ}$ is the corresponding excitation energy, $b^\dagger_{\nu\vq}$ is a creation operator for a phonon with momentum $\vq$ and mode $\nu$, $\omega_{\nu\vq}$ is the phonon frequency, and $\mathcal{G}_{nm\nu}(\vQ,\vq)$ is the exciton-phonon coupling matrix element~\cite{Chen2020,Antonius2022}, which describes the scattering from an exciton in state $(n,\vQ)$ to state $(m,\vQ+\vq)$ via a phonon $(\nu,\vq)$.

The general equation of motion for the exciton density matrix reads,
\begin{equation}
    \frac{d\rho_{nm\vQ}}{dt}+i[H_0,\rho]_{nm\vQ}=\frac{d\rho_{nm\vQ}}{dt}\Bigg|_{col},
    \label{eq:EOM}
\end{equation}
where $\rho_{nm\vQ}=\langle a^\dagger_{n\mathbf{Q}} a_{m\mathbf{Q}} \rangle $ is the exciton density matrix diagonal in the exciton COM $\vQ$, $H_0$ is the quadratic part of the Hamiltonian and the term on the right-hand side is the collision integral---
here it describes the impact of phonon scattering on the evolution of the exciton density matrix.

In this work, we approximate the collision integral using two theoretically well established, but rarely implemented at the \textit{ab initio} level, formulations. The first is derived from a Lindblad superoperator based on the many-body Hamiltonian~\cite{Rosati2014,Iotti2017}, which has been utilized to investigate ultrafast spin dynamics within the electron density matrix formalism~\cite{XuJ2021}.
Hereafter, we refer to this as the ``Lindblad" approach.
For excitons, we can write,
\begin{widetext}
\begin{align}
\frac{d\rho_{ij\mathbf{Q}}}{dt}\Bigg|_{col} & =\frac{2\pi}{\hbar N_\vq}\text{Re}\left\{ \left(I+\rho\right)_{is\mathbf{Q}}\mathcal{\tilde{G}}_{st\nu}^{*}\left(\mathbf{Q},\mathbf{q}\right)\rho_{tm\mathbf{Q}+\mathbf{q}}\mathcal{\tilde{G}}_{jm\nu}\left(\mathbf{Q},\mathbf{q}\right)\right.\nonumber\\
 & \left.-\mathcal{\tilde{G}}_{si\nu}\left(\mathbf{Q}-\mathbf{q},\mathbf{q}\right)\left(I+\rho\right)_{st\mathbf{Q}-\mathbf{q}}\mathcal{\tilde{G}}_{tm\nu}^{*}\left(\mathbf{Q}-\mathbf{q},\mathbf{q}\right)\rho_{mj\mathbf{Q}}\right\} \left(n_{\mathbf{q}}+1\right)\nonumber\\
 & +\frac{2\pi}{\hbar N_\vq}\text{Re}\left\{ \left(I+\rho\right)_{is\mathbf{Q}}\mathcal{\tilde{G}}_{ts\nu}\left(\mathbf{Q}-\mathbf{q},\mathbf{q}\right)\rho_{tm\mathbf{Q}-\mathbf{q}}\mathcal{\tilde{G}}_{mj\nu}^{*}\left(\mathbf{Q}-\mathbf{q},\mathbf{q}\right)\right.\nonumber\\
 & \left.-\mathcal{\tilde{G}}_{is\nu}^{*}\left(\mathbf{Q},\mathbf{q}\right)\left(I+\rho\right)_{st\mathbf{Q}+\mathbf{q}}\mathcal{\tilde{G}}_{mt\nu}\left(\mathbf{Q},\mathbf{q}\right)\rho_{mj\mathbf{Q}}\right\} n_{\mathbf{q}},
  \label{eq:lindblad}
\end{align}
\end{widetext}
where $n_\vq$ is the phonon occupation and we define
\begin{align*}
\tilde{\mathcal{G}}_{jm\nu}(\mathbf{Q},\mathbf{q}) & \equiv\mathcal{G}_{jm\nu}(\mathbf{Q},\mathbf{q})\delta\left(E_{m\mathbf{Q}+\mathbf{q}}-E_{j\mathbf{Q}}-\hbar\omega_{\mathbf{q}\nu}\right)^{1/2},
\end{align*}
and
\begin{align*}
    \mathcal{\tilde{G}}_{im\nu}^{*}\left(\mathbf{Q},\mathbf{q}\right) & \equiv\mathcal{G}_{im\nu}^{*}\left(\mathbf{Q},\mathbf{q}\right)\delta\left(E_{m\mathbf{Q}+\mathbf{q}}-E_{i\mathbf{Q}}+\hbar\omega_{\mathbf{q}\nu}\right)^{1/2}.
\end{align*}
The plus sign in $(I+\rho)$ reflects the Bose statistics and one can show that the Boltzmann equation is recovered by taking only the state diagonal part of the density matrix. 
In deriving Eq.~\ref{eq:lindblad}, a Markovian approximation is adopted, so memory effects are absent in this framework. 
The populations of phonons are assumed to follow a Bose distribution at a fixed temperature, which treats phonons as a bath, ignoring their dynamics. 
Equation~\ref{eq:lindblad} in the semiclassical Boltzmann limit~\cite{Chan2025}, where only the diagonal terms of the collision integral are included, describes the phonon-driven population trasnfers and relaxations.
In contrast, the off-diagonal terms are vital to the decoherence processes. We do not include additional relaxation/dephasing channels in our dynamics.

Although the DP mechanism is traditionally derived using a collision term that is a linear functional of the density matrix, recent work on electron density matrices has shown that the DP regime persists even when nonlinearities are introduced as in the collision integral above~\cite{Xu2020,XuJ2021}. 
We demonstrate in Sec.~\ref{sec:fourstate} that the signature of the DP regime is clearly preserved within our excitonic framework.

The second formulation we investigate derives the collision integral from the NEGF approach, which goes beyond the Markov approximation~\cite{Marini_2013,Perfetto2023}. In general, the derived equation of motion of the density matrix depends on several components of the two-time Green's function. 
To close the equation and avoid solving the full Kadanoff-Baym equation, we make the generalized Kadanoff-Baym ansatz~\cite{Lipavsky1986}. 
Hereafter, we refer to this as the ``GKBA" approach.
Taking into account the Fan diagram of exciton-phonon self-energy~\cite{Chen2020,Antonius2022}, the collision integral reads, $\frac{d\rho}{dt}\Big|_{col}=-\left[I_L^<+H.c.\right]$, where 
\begin{widetext}
\begin{align}
I_{L,ip\vQ}^{<}(t) & =i\sum_{njsm\nu\mathbf{q}}\int_{0}^{t}dt'\mathcal{G}_{\nu in}(\vQ-\mathbf{q},\mathbf{q})\mathcal{G}^*_{\nu jl}(\vQ-\mathbf{q},\mathbf{q})\left[G_{ns\vQ-\mathbf{q}}^{R}(t,t')\bar{\rho}_{sl\mathbf{Q}-\mathbf{q}}(t')D_{\nu\mathbf{q}}^{<}(t',t)\rho_{jm\vQ}(t')G_{mp\vQ}^{A}(t',t)\right.\nonumber\\
 & \left.-G_{ns\vQ-\mathbf{q}}^{R}(t,t')\rho_{sl\vQ-\mathbf{q}}(t')D_{\nu\mathbf{q}}^{>}(t',t)\bar{\rho}_{jm\vQ}(t')G_{mp\vQ}^{A}(t',t)\right].
 \label{eq:GKBA_scat}
\end{align}
\end{widetext}
In Eq.~\ref{eq:GKBA_scat}, $G^R$ and $G^A$ are the retarded and advanced components of the exciton Green's function, respectively; $\bar{\rho}=I+\rho$; and $D^<$ and $D^>$ are the lesser and greater components of phonon propagators, respectively~\cite{QK2008}. In our implementation, we approximate the exciton propagator with the Wigner-Weisskopf form~\cite{Marini_2013,QK2008}, 
\begin{equation}
    G^R_{ns\vQ}(t,t') = -i\theta(t-t')\delta_{ns}e^{-i(E_{n\vQ}-i\eta)(t-t')},
\end{equation}
where $\theta(t)$ is the Heaviside function and $\eta$ is the numerical broadening parameter.
The computational cost of Eq.~\ref{eq:GKBA_scat} scales as $O(t^2)$, posing a significant bottleneck for long-time propagation.
To overcome this, we employ the recently developed G1-G2 scheme, which reformulates the integro-differential equation as a system of coupled first-order ordinary differential equations~\cite{Bonitz2020}.
Specifically, the memory kernel which involves a time-integral of the density matrix is handled by introducing an auxiliary density matrix, the formal derivation of which is provided in the Appendix I A~\ref{sec:g1g2}.

A comparison of these two scattering formulations is instructive.
Both frameworks describe the evolution of band- and momentum-resolved exciton coherences and populations driven by microscopic exciton-phonon scattering events. 
One can show that both collision terms strictly conserve the total exciton population, which we also verify numerically.
In the Lindblad approach (Eq.~\ref{eq:lindblad}), the delta functions enforce energy conservation during exciton-phonon scattering.
In contrast, the GKBA collision integral (Eq.~\ref{eq:GKBA_scat}) allows for off-shell scattering that does not strictly conserve energy at short time scales, consistent with the energy-time uncertainty principle. 
Furthermore, the time-integral formulation preserves the memory effects inherent to the dynamics; notably, this framework has been successfully applied to capture the polaron features in transient absorption spectra~\cite{QK2008}.

\subsection{Helical versus spin basis}

A subtle but fundamentally important issue in modeling dissipative dynamics is basis invariance.
While both Eq. \ref{eq:lindblad} and Eq. \ref{eq:GKBA_scat} account for exciton scattering, the Lindblad approach is not basis independent. 
This stems from the fact that the energy-conserving delta functions are typically defined with respect to a specific set of eigenstates; because these energy arguments do not transform co-variantly under basis changes, the resulting dynamics become sensitive to the representation. As shown in Ref.~\cite{Marini_2013,QK2008}, this basis dependence is a direct consequence of the infinite-time integration implicit in the Markov approximation, which ties the collision operator to the representation in which energy conservation is enforced.

This distinction is particularly relevant in the study of valley or spin dynamics, where two different bases are frequently employed. In the standard derivation of the DP mechanism~\cite{DP1971,Zhu2014,Mower2011,Maialle1993,Hao2016}, excitons with small COM $\vQ$ are expanded in the basis of degenerate $\vQ=0$ A excitons, which are localized in the $K$ or $K'$ valley.
In this representation, the exciton pseudospins are defined by their valley localization. 
The effective Hamiltonian in this basis includes the long-range intervalley exchange interaction-which vanishes at $\vQ=0$---alongside momentum-dependent intravalley exchange and direct Coulomb terms. 
Because the intervalley exchange interaction flips the pseudospin, it acts as an effective transverse magnetic field, driving the precession dynamics.
We denote the corresponding formulation of the collisions as the ``Lindblad spin" approach.

Alternatively, in Eq.~\ref{eq:Ham}, exciton states are obtained by solving the first-principles Bethe-Salpeter equation (BSE), which incorporates both direct and exchange Coulomb kernels.
We refer to the resulting eigenstates as the helical basis and denote the corresponding treatment as the ``Lindblad helical" approach.
In the following sections, we demonstrate that numerical results obtained from these two bases are indeed different. 
In stark contrast, Eq.~\ref{eq:GKBA_scat} and Eq.~\ref{eq:EOM} are co-variant under basis transformations, ensuring that physical observables remain strictly invariant.

\subsection{Computational details}

We have implemented a numerical solver for both Lindblad and GKBA approaches. 
All of the required ingredients, such as phonon modes, exciton energies, and exciton-phonon coupling matrix elements, are computed from first-principles. 
The time-propagation is integrated with a fourth order Runge-Kutta algorithm with a step size of 0.5 fs.
Four exciton bands are considered in the calculations. 
Numerically, delta functions are implemented as Lorentzian functions with a width of 20 meV, which is equivalent to $\eta=10$ meV in the Wigner-Weisskopf exciton propagator.

Computational details of the first-principles calculations are given as follows. Electronic structure and density functional perturbation theory~\cite{Baroni2001} calculations are performed with the Quantum Espresso package~\cite{qtespresso}.
Electron-phonon coupling matrix elements are computed with the EPW package~\cite{PONCE2016}.
We compute the GW quasi-particle energy and solve for excitons with the BerkeleyGW package~\cite{Hybertsen1986,Rohlfing2000,Deslippe2012}. 
Finite COM excitons~\cite{Qiu2015} and phonons are solved on a $48\times48$ $\vQ$-grid and $\vq$-grid, respectively, on which exciton-phonon coupling matrix elements are computed.
All other parameters can be found in our previous work~\cite{Chan2023}.
\begin{figure}[t]
     \centering
     \includegraphics[width=0.48\textwidth]{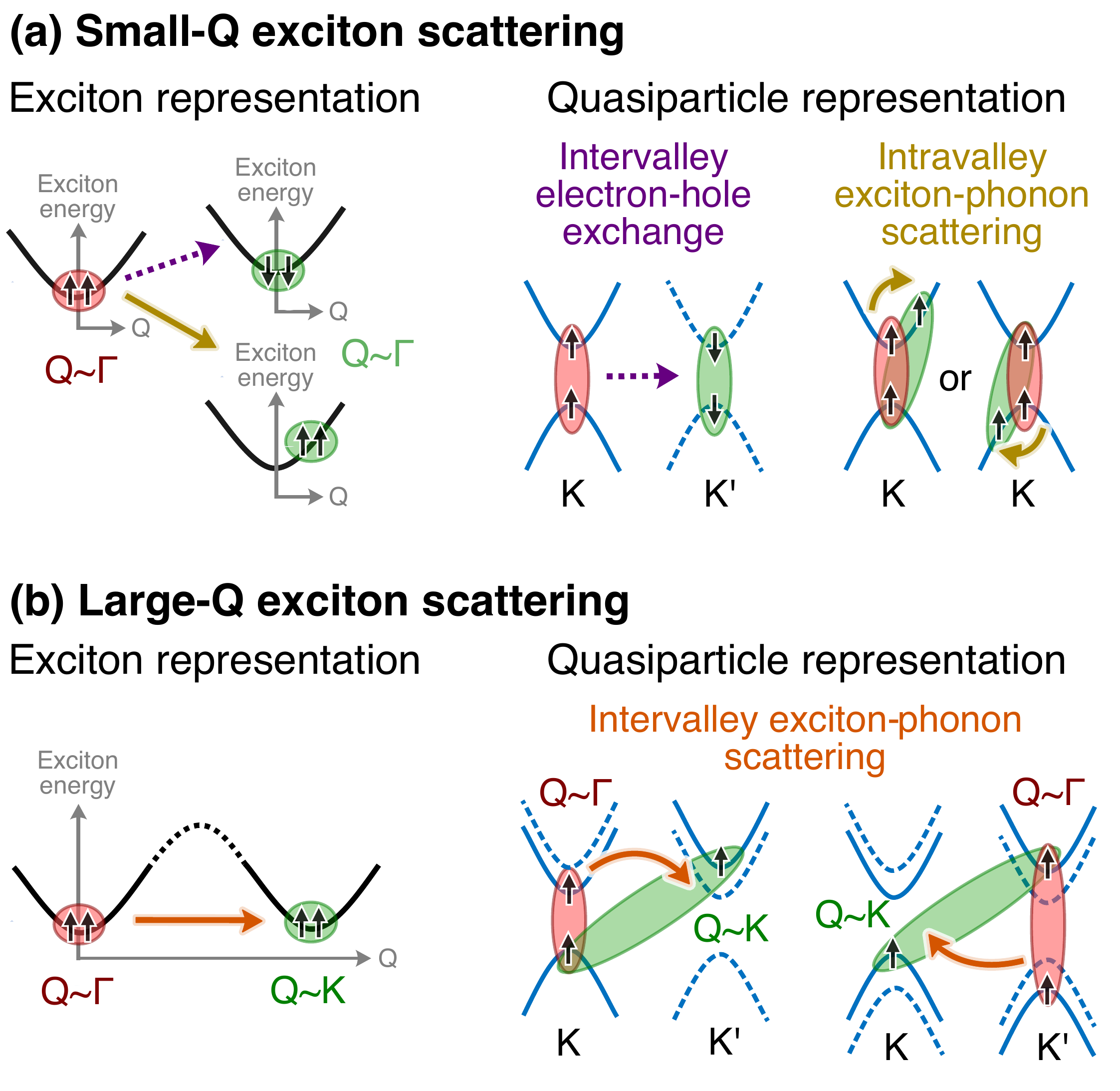}
\caption{(a) Schematic representation of scattering processes associated with small transfer of exciton center-of-mass (COM) momentum $\mathbf{Q}$, and visualized both in the exciton and quasiparticle band structures. Small-$\mathbf{Q}$ exciton scattering processes originate from both intervalley electron-hole exchange interactions, which strictly conserve $\mathbf{Q}$, and intravalley exciton-phonon scattering processes, which can be broken down into electron-phonon and hole-phonon scattering contributions. (b) Corresponding representation of processes associated with a large transfer of the exciton COM momentum $\mathbf{Q}$. For large-$\mathbf{Q}$ exciton scattering, the primary mechanism is intervalley exciton-phonon scattering processes, which is composed of electron- and hole-phonon scattering contributions.
}
\label{fig:valley_coupl}
\end{figure}

\begin{figure*}[t]
     \centering
     \includegraphics[width=\textwidth]{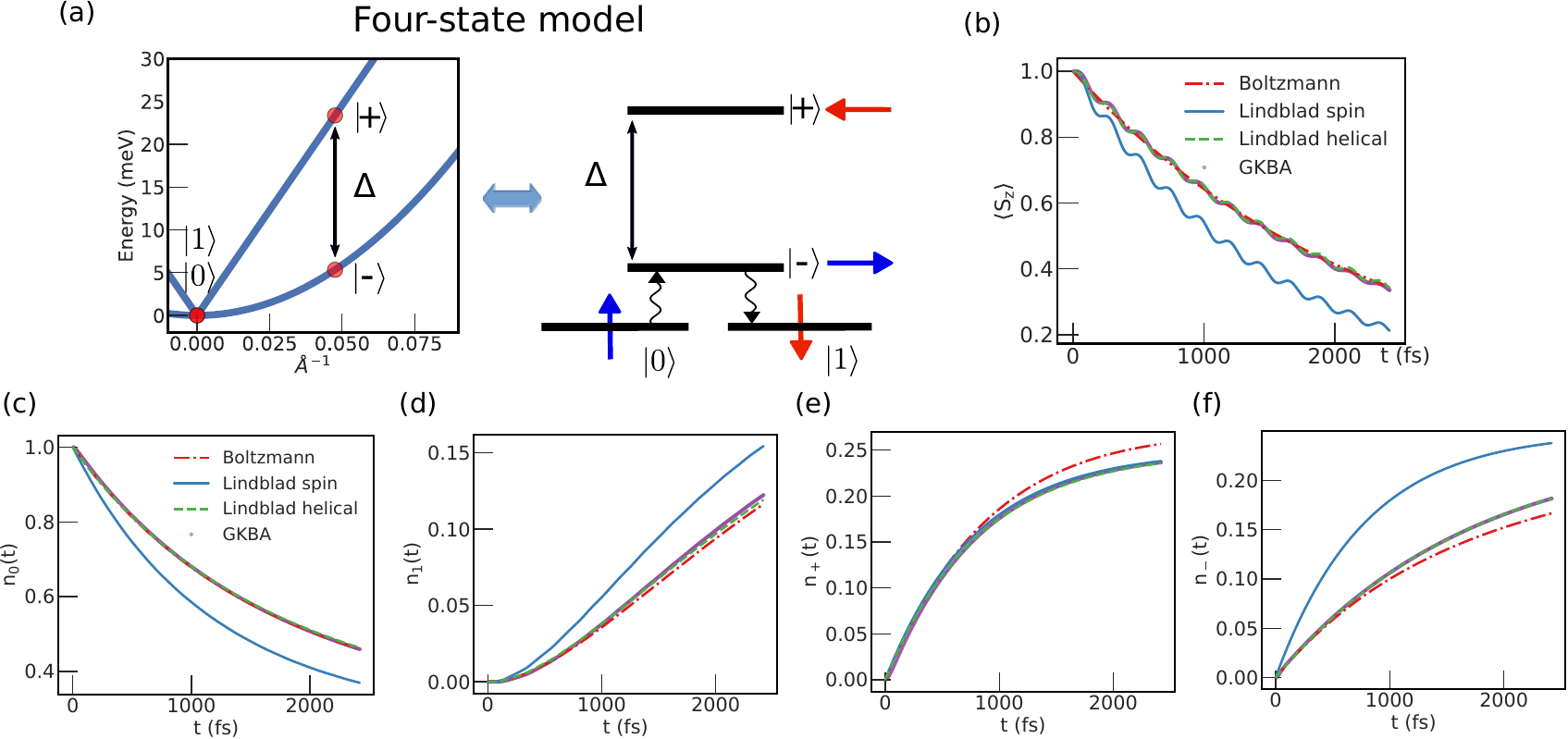}
\caption{(a) Four exciton states selected from the exciton bands define the four-state model. $\ket{0}$ and $\ket{1}$ states are degenerate A excitons. $\ket{+}$ and $\ket{-}$ are exchange-split states at $\vQ=(1/48,0)$. (b) Time evolutions of valley polarizations $\langle S_z\rangle$ from the four approaches. (c)-(f) Time evolutions of populations of the four states, respectively. The results shown are simulated by Boltzmann (red dot-dashed line), Lindblad spin (blue solid line), Lindblad helical (green dashed line), and GKBA (purple dots) approach. 
}
\label{fig:fourstate}
\end{figure*}

\section{Intravalley exciton dynamics}
\label{sec:intravalley}

The DP mechanism for small momentum excitons has previously been adopted to interpret experiments~\cite{Zhu2014,Hao2016}.
Indeed, a minimum four-state model~\cite{Park2022} clearly demonstrated the crossover of valley dynamics between the DP and other regimes depending on the magnitude of the momentum scattering rate, an approach we follow here to verify its applicability to our system.
However, the role of large-momentum transfer scatterings, which leads to ill-defined valley quantum numbers, remains poorly understood.
To disentangle the influence of various relaxation channels, we distinguish between scattering processes associated with small and large transfer of exciton momentum Q.
Small-Q scattering processes include intravalley exciton-phonon scattering, wherein both the electron or the hole acquire a small momentum change, or exchange intervalley interaction, wherein both electrons and holes flip the valley, but the overall exciton momentum remains conserved [Fig.~\ref{fig:valley_coupl}~(a)]. 
In contrast, only exciton-phonon interactions lead to scattering processes that contribute to a large change in the exciton momentum Q [Fig.~\ref{fig:valley_coupl} ~(b)].
In this section, by focusing first on the small-momentum ($\mathbf{Q} \approx 0$) exciton dynamics, we investigate the competition between these exchange-driven mechanisms and phonon-induced scattering, thereby identifying the hallmarks of the DP mechanism.

\subsection{Four-state model}
\label{sec:fourstate}
We start with a four-state model, where only A excitons and two excitons of $\vQ=(1/48,0)$ (The smallest $Q$ commensurate with our uniform grid) are included as shown in Fig.~\ref{fig:fourstate} (a).
The quadratic part of the Hamiltonian in the spin basis is
\begin{align}
    H_{spin}=\frac{\epsilon_2}{2}(\sigma_z+I)\otimes I+\frac{\Delta}{4}(\sigma_z+I)\otimes\tau_x,
    \label{eq:Hspin}
\end{align}
where $\epsilon_2$ is the energy of exciton at $\vQ=(1/48,0)$ without exchange interaction offset by the $\vQ=0$ exciton energy, $\sigma_z$ is the Pauli $z$-matrix in the space of COM $\vQ=0$ and $\vQ=(1/48,0)$ and $\tau_x$ is the Pauli $x$-matrix in the space of degenerate $K$ and $K'$ valleys.
The two states with finite momentum in the helical basis, obtained by diagonalizing Eq.~\ref{eq:Hspin}, are denoted as $\ket{+}$ and $\ket{-}$, respectively.
Their energy splitting $\Delta$ is due to intervalley exchange interactions.
Exciton-phonon interactions couple the A exciton states ($\ket{0}$ and $\ket{1}$) to the $\ket{+}$ and $\ket{-}$ states but couplings between orthogonal spin states are not possible since exciton-phonon interactions do not flip pseudospin in this energy range. The exciton energy and coupling strength are taken from our first-principles calculations (Sec. II. C).

To model photo-excitation by circularly polarized light, we initialize our simulation in state $\ket{0}$, corresponding to a valley polarized configuration with $\langle S_z\rangle=1$ and monitor the subsequent evolution of populations and valley polarizations.
The temperature of the phonon bath is set at 300 K.
In the following, the $\langle S_z\rangle$ is computed by summing the contributions of the zone center exciton and its six nearest neighbors.

Figures~\ref{fig:fourstate}(c)–(f) illustrate the population dynamics for the four-state system calculated with our four numerical approaches. 
In the Boltzmann limit, Eq.~\ref{eq:EOM} and \ref{eq:lindblad} are reduced to their diagonal components, thereby neglecting exciton coherences. 
Qualitatively, all four methodologies yield consistent dynamical trends. 
Quantitatively, the results from the Boltzmann, Lindblad-helical, and GKBA approaches are in close agreement. 
In each case, an initial transition from the $\ket{0}$ state to the higher-energy $\ket{+}$ and $\ket{-}$ states, manifesting as a rapid population surge in the latter, is followed by the decay of these high-energy states into the $\ket{1}$ state. 
The convergence of the three approaches suggests that the helical basis effectively diagonalizes the dominant interactions.

\begin{figure*}[t]
     \centering
     \includegraphics[width=.96\textwidth]{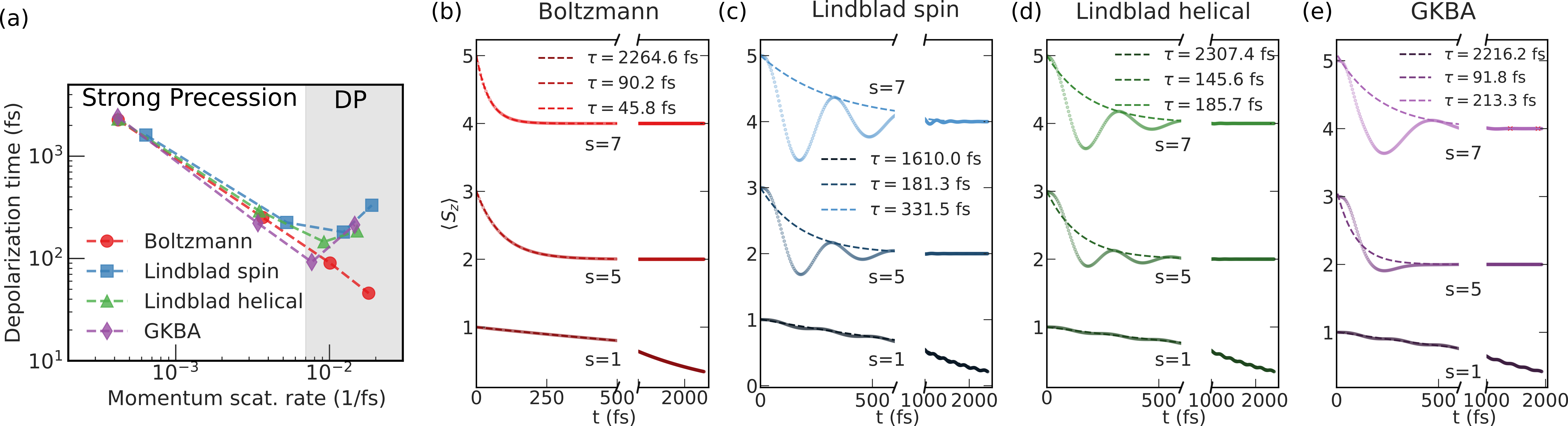}
\caption{(a) Depolarization time as a function of scattering rate from the simulations of the four approaches. The shaded region indicates the DP regime, which is characterized by the upturn of the curves from the three simulations other than the Boltzmann approach. (b)-(e) Valley polarizations dynamics simulated by the Boltzmann, Lindblad spin, Lindblad helical, and GKBA approach, respectively. Curves obtained from simulations with $s\neq1$ are vertically offset for clarity. In each panel, the dots are simulated data and dashed lines are the best fitting exponential functions. Curves with amplitude oscillations are fitted by their local maximums. 
}
\label{fig:fourstate_DP}
\end{figure*}

Figure ~\ref{fig:fourstate} (b) shows the time evolution of the valley polarization.
While the four approaches yield qualitatively consistent trends with the Boltzmann, Lindblad-helical, and GKBA providing quantitatively close results, the Boltzmann approach fundamentally fails to capture the coherent oscillations dynamics in the valley polarization. 
Instead, it shows a mono-exponential decay, confirming that a population-based description is insufficient to resolve the quantum coherence of valley dynamics.

The coherent oscillation can be understood as the precession of pseudospin around an effective transverse field, with the angular frequency determined by the energy splitting $\Delta$.
This precession is most intuitively described in the spin basis, where superposition of up and down valley states is encoded in the off-diagonal terms of the density matrix.
Because these coherences are completely neglected in the Boltzmann framework, the precession can not be captured, resulting in the loss of the oscillating polarizations.
Notably, the populations calculated using the Lindblad spin approach equilibriate more rapidly compared to the other three methods, all of which remain in close agreement.
These results provide direct numerical evidence of the basis dependence inherent to the Lindblad formulation as discussed in Section~\ref{sec:method} B.

We emphasize that the polarization relaxation in this model is a combined result of exchange interactions and exciton-phonon scatterings.
Without exchange interactions, valley spin $S_z$ is a conserved quantity while turning off the exction-phonon scatterings leads to either frozen dynamics or persistent precession depending on the initial state.

As demonstrated in Ref.~\cite{Park2022}, the four-state model exhibits a dynamical crossover spanning three distinct regimes-Elliot-Yaffet (EY), DP, and strong precession-determined by the relative strength between the momentum scattering rate, the exchange splitting and the spin-flip scattering rate.
Since exciton-phonon interactions do not introduce valley-pseudospin flips in this model, the EY mechanism is inactive, and we focus on the latter two regimes.
Unlike the DP regime discussed in Sec.~I, the strong precession regime is characterized by fast precession dynamics such that the time-averaged spin aligns with the precession axis; consequently, polarization relaxes through the realignment of this axis.
This leads to a fundamental change in the scaling behavior.
While the relaxation rate is inversely proportional to the scattering rate in the DP regime (motional narrowing), it becomes directly proportional to the scattering rate in the strong precession limit.
In the following, we resolve this crossover by systematically scaling the exciton-phonon coupling strength by a factor $s$.

\begin{figure*}[t]
     \centering
     \includegraphics[width=\textwidth]{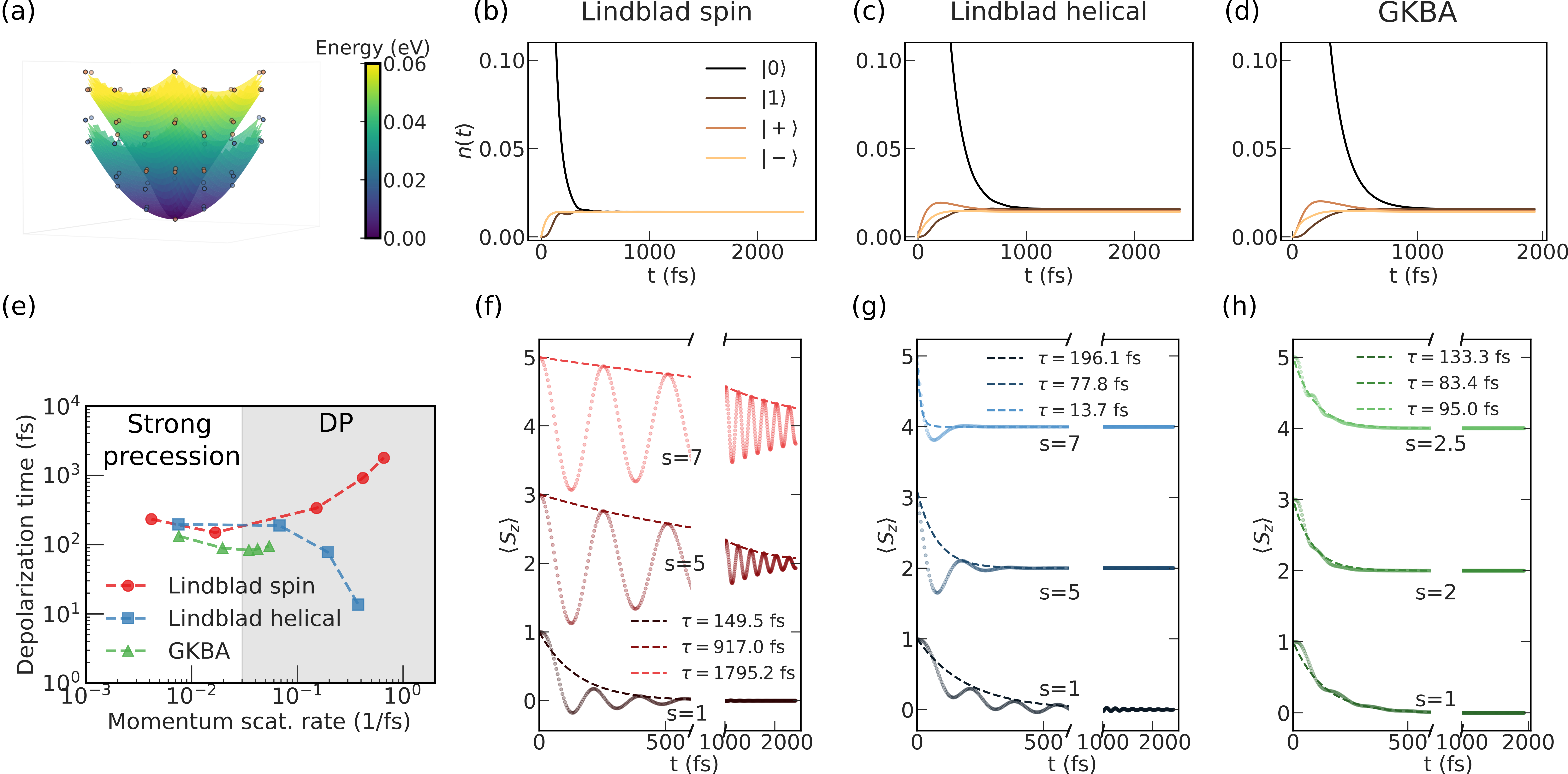}
\caption{(a) Two dimensional exciton energy dispersion near center-of-mass momentum, $\vQ=0$. Red dots are exciton states included in the calculation restricted to small $\vQ$. (b)-(d) are exciton populations dynamics of the four states defined in Fig. 1, from Lindblad spin, Lindblad helical, and GKBA approach, respectively. (e) Depolarization time as a function of scattering rate from the simulations of the three approaches. Shaded region indicates the DP regime for the Lindblad spin and GKBA simulations. (f)-(h) Valley polarization dynamics with scaled exciton-phonon couplings from the three approaches. See the caption of Fig.~2 for definitions of the curves and symbols.
}
\label{fig:patch}
\end{figure*}

Figures~\ref{fig:fourstate_DP} (b)-(e) display the polarization dynamics from four different approaches with $s=1$, 5, and 7, with the $s\neq1$ curves vertically offset for clarity.
At $s=1$, the relaxation rates extracted from the Boltzmann, Lindblad helical, and GKBA simulations are in quantitative agreement.
However, as $s$ increases, the results diverge significantly.
The Boltzmann approach yields a relaxation time that monotonically decreases with $s$. 
In contrast, the other three frameworks exhibit a transition to underdamped oscillatory behavior ($\Delta\tau/\hbar>1$, where $\tau$ is the depolarization time).
By fitting the relaxation time from the decay envelopes of these oscillations, we find that it exhibits a non-monotonic dependence on $s$; it initially decreases before increasing in tandem with the oscillation amplitude. 
This 'upturn' in depolarization time is the definitive hallmark of the DP mechanism, which underscores the fundamental importance of exciton coherence in describing valley dynamics.

While the precession rates extracted from the two Lindblad approaches are consistent with each other, the GKBA yields a smaller rate, which might be attributed to the higher order vertex corrections embedded in the GKBA collision integral.
A comparison between the two Lindblad implementations reveals a stronger precession in the spin basis. This can be understood by noting that the $\ket{+}$ and $\ket{-}$ states in the helical representation have an energy offset due to exchange, resulting in an imbalanced population across the basis states that effectively suppresses the coherent precession amplitude.

The transition between these regimes is summarized in Fig.~\ref{fig:fourstate_DP} (a), where the depolarization time is plotted against the momentum scattering rate extracted from early-time exciton populations.
The clear upturn signature at high scattering rates confirms the crossover into the DP regime—a feature entirely absent in the Boltzmann results. These findings demonstrate that, while the predicted relaxation times vary, the exciton density matrix approach (both Lindblad and GKBA) is capable of capturing the complex interplay between coherence and dissipation that defines the DP mechanism.

\subsection{\textit{Ab initio} calculation restricted to small $Q$}

\begin{figure*}[t]
     \centering
     \includegraphics[width=\textwidth]{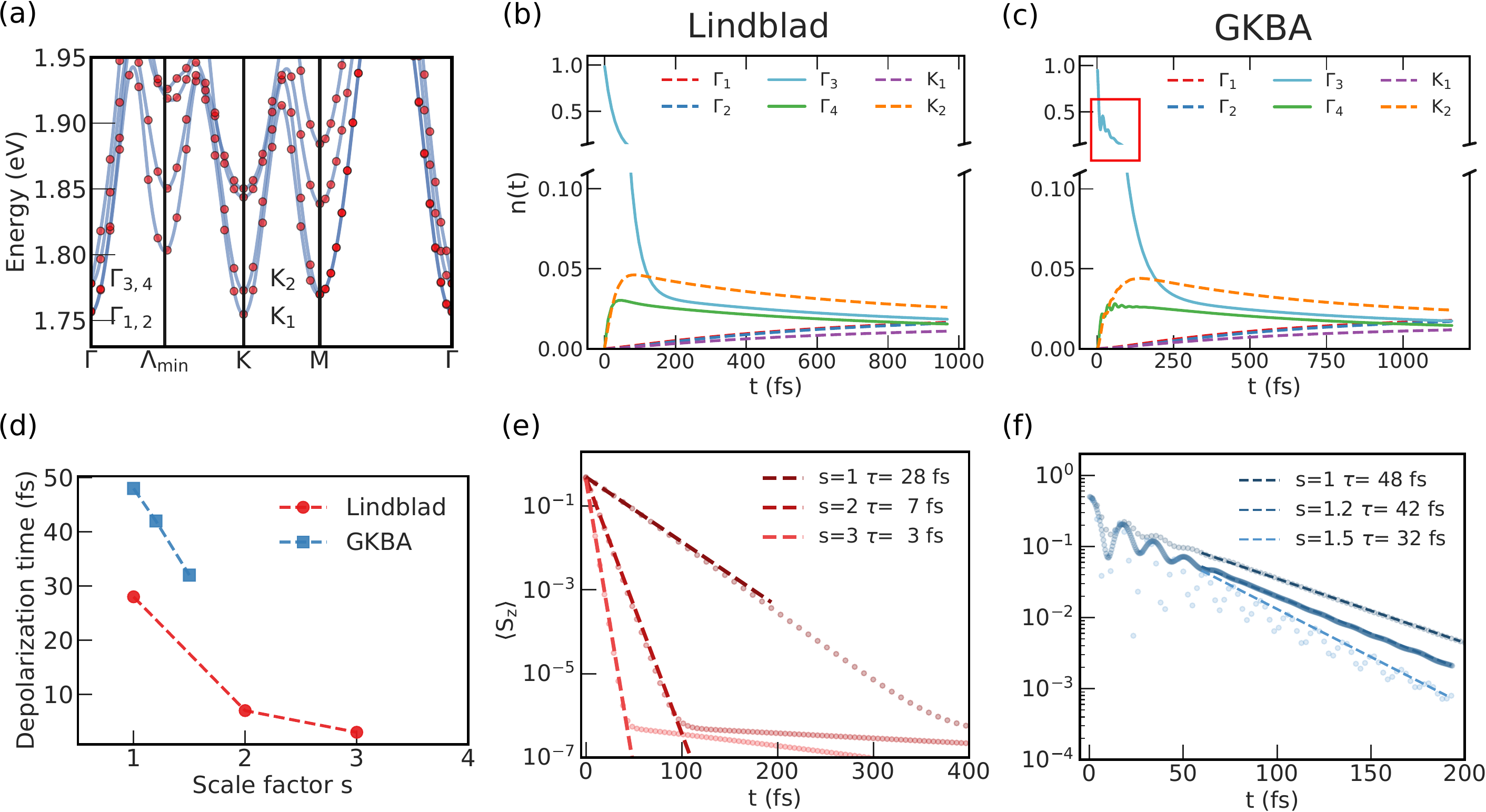}
\caption{(a) Exciton energy dispersion along the high symmetry path. States at $\Gamma$ and $K$ are labeled by their energy orders at that valley.
(b) and (c) show exciton population dynamics of six low energy exciton states from the Lindblad helical and GKBA simulations. Dashed lines indicate states which are optically dark. Red box in (c) highlights the transient oscillations. (d) Depolarization time as a function of exciton-phonon couplings scale factors. (e) and (f) show valley polarization dynamics from the two approaches, respectively. Symbols are simulated data while dashed lines are the best fitting function in the same time span.}
\label{fig:fullBZ}
\end{figure*}

Next, we extend our analysis to the dynamics of excitons with COM $\vQ$ in a 0.15 $\AA^{-1}$ radius patch centered at $\vQ=0$ as illustrated in Fig.~\ref{fig:patch} (a).
The exciton energy, exchange couplings, and exciton-phonon coupling strength are taken from our $\textit{ab initio}$ calculations (Sec. II.C).
Having established that the Boltzmann formulation is not capable of capturing the DP mechanism, we focus on the simulations using Lindblad and GKBA approaches.
Figures~\ref{fig:patch} (b)-(d) display the population dynamics for the four characteristic states within this patch.
Qualitatively, all three simulations exhibit consistent trends of the population evolution.
However, the Lindblad spin approach predicts a faster population decay rate, with the A exciton population reaching equilibrium at approximately 500 fs, nearly twice as fast as the 1 ps scale predicted by the other two approaches. 
Quantitatively, Lindblad helical and GKBA results converge, suggesting that the helical representation, as opposed to spin representation, may be a more robust basis for the dissipative dynamics in this regime.

Figures~\ref{fig:patch} (f)-(h) display valley polarization dynamics for several scaled exciton-phonon coupling strengths, revealing divergent behaviors across the three simulations.
The Lindblad spin approach predicts underdamped oscillations with a precession time of 200 fs.
In stark contrast, the other two methods exhibit heavily damped dynamics, with the GKBA specifically yielding a vanishing precession amplitude.
As we increase the scaling factor $s$, the Lindblad spin results show enhanced precession amplitudes in tandem with the depolarization time, which indicates that the system resides in the DP regime even at $s=1$.
Conversely, the Lindblad helical simulation shown in Fig.~\ref{fig:patch} (g) displays a monotonically decreasing depolarization time with increased coupling, suggesting the absence of motional narrowing.
Notably, we no longer observe the coherent precession in the strongest coupling case, which can be rationalized by the fact that the scattering phase space in the calculation restricted to small $\vQ$ is much larger and the previously discussed population imbalances suppress the coherent oscillation.

In the GKBA simulation, we observe a non-monotonic trend of the depolarization time: it initially decreases but increases subsequently at higher coupling strength as shown in Fig.~\ref{fig:patch} (h).
While simulations at the high coupling values suffer from known numerical instabilities which can lead to negative exciton populations~\cite{Gianluca2024} preventing us from scaling to $s>3$, the results at $s=2.5$ clearly show a longer relaxation time than at $s=2$. 
This indicates a crossover into the DP regime at a modest coupling strength.

The transition is summarized in Fig.~\ref{fig:patch} (e).
While the crossover between the DP and the strong precession regime is captured by both the Lindblad spin and the GKBA simulations, their predicted relaxation times differ.
The Lindblad spin results align with previous model-based studies~\cite{Zhu2014,Hao2016} that attribute the depolarization time to the DP mechanism.
However, our simulation with the advanced GKBA treatment suggests that for a single exciton valley, the intravalley exciton-phonon coupling is too weak to drive DP physics so that the system remains in the strong precession regime.

\section{\textit{Ab initio} calculations in the full BZ}
\label{sec:fullBZ}

Having analyzed the competition between exciton-phonon scattering and exchange interactions within a single exciton valley, we extend our analysis to the entire BZ, including all low-energy dark exciton branches illustrated in Fig.~\ref{fig:fullBZ} (a).

Figures~\ref{fig:fullBZ} (b) and (c) display the population dynamics of characteristic exciton states, integrated over 0.15 $\AA^{-1}$ patches centered at $\vQ$ points labeled in (a), as calculated using the Lindblad helical and GKBA frameworks at 300 K, respectively.
Consistent with our previous results, the two approaches yield qualitatively similar trends of population dynamics.
The population of $\Gamma_3$ (A) exciton exhibits a rapid decay within the first 250 fs before asymptotically approaching equilibrium.
While the populations of $\Gamma_4$ (the other A) and $K_2$ exciton surge, overshooting their equilibrium values before slow subsequent decay.
Crucially, the rapid rise of the $K_2$ population within the first 100 fs highlights the critical role of large-momentum intervalley exciton-phonon scattering as the dominant valley depolarization channel---a feature inherently absent in our previous phase space-restricted small $\vQ$ patch.
The Lindblad approach predicts a slightly accelerated population dynamics, with the $\Gamma_4$ and $\Gamma_3$ exciton populations equalizing around 200 fs compared to 250 fs in the GKBA. More striking, the GKBA results exhibit non-Markovian transients, not seen in the Lindblad simulation.
Specifically, in Fig.~\ref{fig:fullBZ} (c), we observe a sharp drop in the A exciton population accompanied by high-frequency oscillations within the first 50 fs.
As this frequency does not correspond to characteristic phonon energies or exchange splittings, and are mirrored in the dynamics of the auxiliary density matrices, they signal a non-Markovian effect, mediated by virtual transition to the high-energy exciton states captured by the GKBA memory kernel, to be discussed in greater detail below.

To determine the accessibility of the DP regime for realistic simulations that include all states in the full BZ, we again perform simulations with scaled exciton-phonon couplings $s$.
Figures~\ref{fig:fullBZ} (e) and (f) depict the resulting valley polarization dynamics. 
We find that the short-time valley polarization dynamics is well-described by a mono-exponential decay.
In the GKBA simulations, the high-frequency oscillations of the polarization within the first 50 fs are directly correlated with the $\Gamma_3$ exciton population transients shown in Fig.~\ref{fig:fullBZ} (c). 
The amplitude of these features increases with the scale factor $s$ until the dynamics becomes numerically unstable at $s=1.5$, where the positivity of the population is no longer preserved.
Notably, the depolarization time in both frameworks decreases monotonically with increasing coupling strength, showing no signatures of the motional narrowing hallmark of the DP mechanism.
Moreover, exciton coherence, which is essential for precession dynamics, remains negligible across all tested coupling strengths. 
Figure~\ref{fig:fullBZ} summarizes the dependence of the depolarization time on the scaling factor $s$.
We therefore conclude that the system remains in the strong precession regime, with intervalley scattering effectively suppressing the coherent DP pathway. This is further evidenced by a significant reduction in the valley depolarization time, which is 3-4 times smaller than the values obtained in the calculation restricted to small $\vQ$, across the different approach employed.
A similar reduction in the valley depolarization time when including the intervalley scattering is also found in our calculations at 10 K, even though the direct scattering rate to COM $K$ exciton is suppressed at low temperature.

\begin{figure*}[t]
     \centering
     \includegraphics[width=\textwidth]{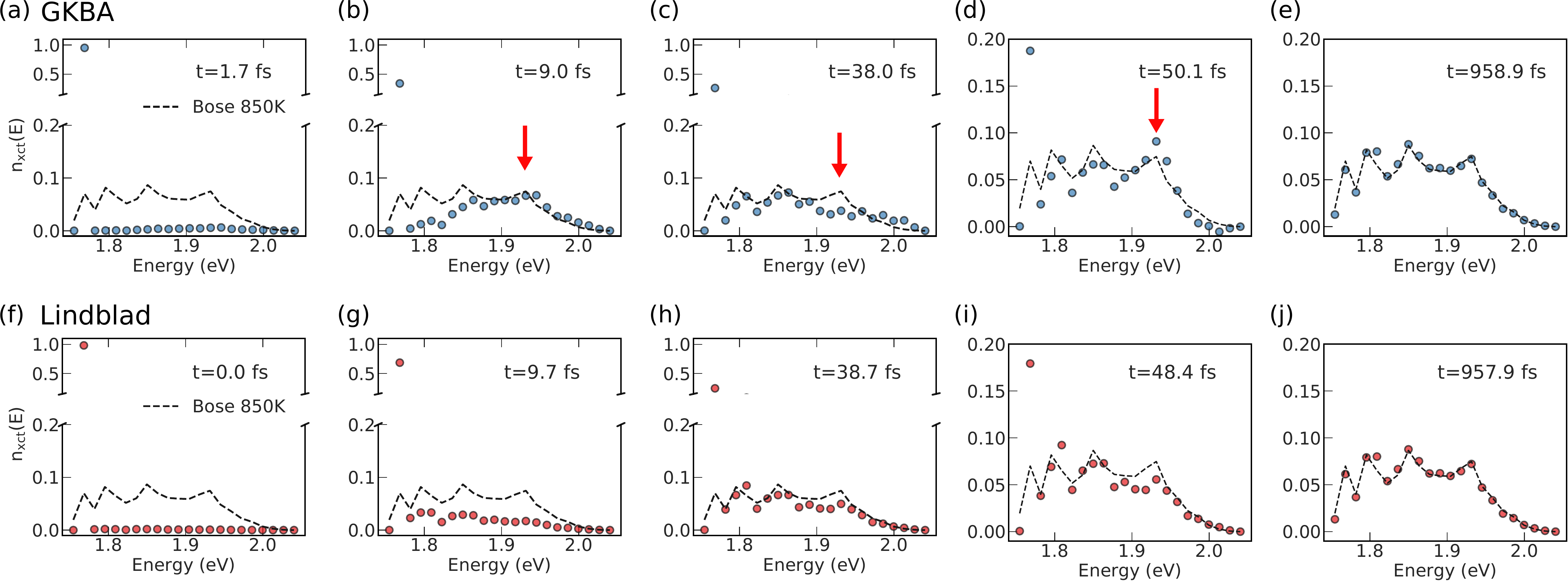}
\caption{(a)-(e) and (f)-(j) display snapshots of exciton distribution functions obtained from GKBA and Lindblad dynamics, respectively. Red arrows in panel (b)-(d) highlight the oscillations of high energy populations. Symbols are data from the simulation. Dashed black lines are Bose distribution at 850 K.}
\label{fig:pop_snapshots}
\end{figure*}

The non-Markovian effect captured by the GKBA simulation is elucidated by further analyzing the evolution of exciton energy distribution functions during the initial stage of relaxation. 
Figures~\ref{fig:pop_snapshots} (a) and (f) depict the initial state for both frameworks, characterized by a single A exciton population.
Although the low-energy exciton population dynamics appear qualitatively consistent trend (cf. Fig.~\ref{fig:fullBZ} (b) and (c)), 
a striking difference emerges in the high-energy exciton distribution within 10 fs as shown Fig.~\ref{fig:pop_snapshots} (b) and (g).
In the GKBA simulation, we observe a significant transient population of high-energy excitons near 1.95 eV. 
In contrast, most exciton populations remain near the A exciton, with a small tail extending toward high energy in the Lindblad calculation.
This distinction arises from the fundamental treatment of scattering: while the Lindblad framework imposes strict energy conservation constraint on the scattering phase space, the GKBA naturally incorporates off-shell processes over short timescales-governed by the energy-time uncertainty principle-allowing for transitions that violate energy conservation at $t\to0$.

By 40 fs, cf. Fig.~\ref{fig:pop_snapshots} (h), high-energy states begin to populate in the Lindblad calculations.
Conversely, the GKBA population exhibits a rapid redistribution back toward low-energy states as shown in Fig.~\ref{fig:pop_snapshots} (c)

At 50 fs, while the overall distribution remains steady in the Lindblad simulation, a dramatic oscillatory resurgence at 1.95 eV is observed in the GKBA distribution.
The rapid oscillations in the high-energy exciton population match those observed in the A exciton population in Fig.~\ref{fig:fullBZ} (c).

These high-frequency oscillations in the high-energy population precisely match the transients observed in the $A$-exciton and polarization dynamics (Fig.~\ref{fig:fullBZ}(c) and (f)), confirming their origin as non-Markovian memory effects unique to the GKBA.

In the long-time limit ($t\approx1$ ps), both distributions nearly converge to an identical high-temperature Bose-Einstein distribution at 850 K.
While a Bose-Einstein distribution is a formal steady state solution for both frameworks, the elevated final temperature is a direct consequence of broadening in the numerical approximation of the energy-conserving delta function (or equivalently, the imaginary part of the self-energy used in the Wigner-Weisskopf propagator).
We find that the effective final temperature can be systematically reduced by employing a smaller broadening parameter.
The ultrafast non-Markovian transient predicted by the GKBA approach may be identified by attosecond spectroscopy~\cite{Lucchini2021,Schumacher2023}.

\begin{figure}[t]
     \centering
     \includegraphics[width=0.45\textwidth]{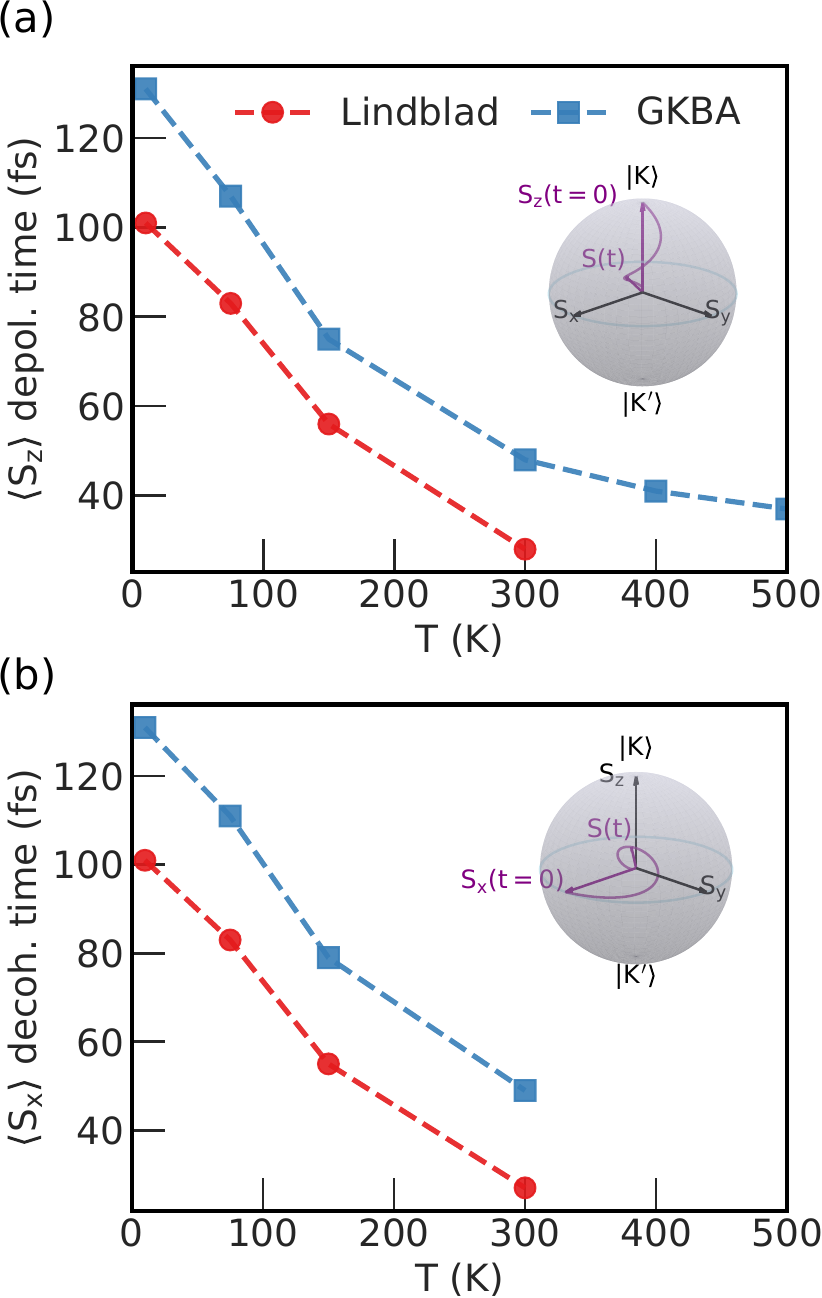}
\caption{(a) Depolarization and (b) decoherence time as a function of temperature fitted from Lindblad (red dots) and GKBA (blue squares) simulations.}
\label{fig:Tdep}
\end{figure}

Valley depolarization and decoherence time of monolayer transition metal dichalcogenides have been extensively characterized through various experimental techniques.
To facilitate a direct comparison with experiments, we present the temperature dependence of our calculated depolarization and decoherence times in Fig.~\ref{fig:Tdep} (a) and (b), respectively.
For decoherence simulations, we initialize the system in a pure state which is an equal superposition of the two degenerate A excitons, which corresponds to excitations by linearly polarized light, and monitor the subsequent evolution of the pseudospin expectation value, $\langle S_x(t)\rangle$.
Detailed simulation data and the corresponding fittings are provided in the Appendix II.
Overall, the GKBA framework yields an approximately 20-30 fs longer valley depolarization and decoherence time than those obtained via the Lindblad approach.
At 300 K, we calculate a valley depolarization and decoherence time of 50 fs, which increase monotonically to 130 fs at 10 K.
These results are in agreement with an earlier four-wave mixing experiment that reported a valley decoherence time of 100 fs at 10 K~\cite{Hao2016} in monolayer WSe$_2$ as well as recent helicity-resolved transient absorption measurements showing a valley decoherence time of 100 fs in the initial stage of the dynamics~\cite{Dogadov2026}.
While our simulations align with these two ultrafast measurements, we note that the latter experiment exhibits a weaker temperature dependence than our predictions and both measurements are done in monolayer WSe$_2$ rather than MoS$_2$.
Furthermore, we acknowledge that several earlier studies reported depolarization times on the order of picoseconds.
These discrepancies may arise from differences in sample quality, exciton densities, or the specific temporal resolution by different spectroscopic techniques.

\section{Conclusion}
\label{sec:conclusion}

In conclusion, we have employed a first-principles exciton density matrix approach, incorporating a non-Markovian collision integral within the GKBA framework, to investigate valley dynamics in monolayer MoS$_2$.
Our findings demonstrate that the DP mechanism is not responsible for valley relaxation in this material owing to insufficient small-momentum exciton-phonon scattering.
Instead, large-momentum intervalley scattering serves as the primary channel for relaxation. This provides parameter-free microscopic support for the intervalley-scattering channel identified in recent transient circular dichroism experiments~\cite{Dogadov2026}, and explains why model analyses restricted to small-momentum excitons have instead inferred DP behavior.

By rigorously accounting for Coulomb-mediated interactions, including both direct and exchange terms, alongside exciton-phonon scatterings, our method allows for an unambiguous evaluation of these couplings on the dynamics.
Equally importantly, we have shown that the widely used Lindblad collision integral is basis-dependent---a direct consequence of its infinite-time Markov limit--and that, in the valley-pseudospin basis underlying prior model analyses, it produces a spurious motional-narrowing signature. Basis-covariant treatments such as the GKBA are therefore essential for reliably identifying relaxation regimes
Furthermore, we identify signatures of non-Markovian effects unique to the GKBA framework, which manifest as rapid transient oscillations during the initial stage of the dynamics.
The calculated valley decoherence times show good agreement with existing four-wave mixing and transient absorption experiments.
Our work clarifies the fundamental relaxation channel of the valley degree of freedom and establishes the GKBA-based density matrix approach as a predictive tool for studying ultrafast exciton dynamics in quantum materials.

\section*{Acknowledgments}
\label{sec:acknowledgments}
This work was supported by the Center for Computational Study of Excited State Phenomena in Energy Materials (C2SEPEM), which is funded by the U.S. Department of Energy, Office of Science, Basic Energy Sciences, Materials Sciences and Engineering Division under Contract No. DE-AC02-05CH11231, as part of the Computational Materials Sciences Program.
F.H.J. was supported by the U.S. Department of Energy (DOE), Office of Science, Office of Basic Energy Sciences (BES), under award DE-SC0021984.
We acknowledge the use of computational resources at the National Energy Research Scientific Computing Center (NERSC), a DOE Office of Science User Facility supported by the Office of Science of the U.S. Department of Energy under Contract No. DE-AC02-05CH11231. The authors acknowledge the Texas Advanced Computing Center (TACC) at The University of Texas at Austin for providing HPC resources that have contributed to the research results reported within this paper. YHC acknowledges the support
of Academia Sinica under project No. AS-CDA-114-M04
and thanks the National Center for High-Performance Computing (NCHC) for providing computational and storage resources.

\section*{Appendix I: Derivation of the GKBA collision integral}
\label{sec:appendix1}

In this appendix, we give the derivation of the collision term with GKBA and the details of the G1-G2 scheme. 
Our derivation closely follows those in Ref.~\cite{Marini_2013} and Ref.~\cite{NEQ2013}.
We start with the equation of motion of the density matrix, which is derived by taking the time-diagonal part of the lesser Green's function in the Kadanoff-Baym equation (KBE),
\[
\frac{d\rho_{ij\mathbf{Q}}}{dt}+i\left[H(t),\rho(t)\right]_{ij\mathbf{Q}}=-\left[I_{L,ij\mathbf{Q}}^{<}+H.c.\right],
\]
where $H(t)$ and $I^<_L$ are defined in the main text.
The collision integral reads,
\begin{align*}
I_{L}^{<}(t) =\int_{0}^{t}dt'\left[\Sigma^{>}(t,t')G^{<}(t',t)-\Sigma^{<}(t,t')G^{>}(t',t)\right],
\end{align*}
where $\Sigma^{>,<}$ and $G^{>,<}$ are the greater and lesser components of self-energy and Green's function, respectively.
We consider a Fan term-like self energy for exciton-phonon interactions~\cite{Chen2020,Antonius2022},
\begin{align}
\Sigma_{ij\vQ}^>(t,t')=&i\sum_{\nu,n,l,\vq}\mathcal{G}_{\nu in}(\vQ-\vq,\vq)\mathcal{G}^*_{\nu jl}(\vQ-\vq,\vq)\nonumber\\
&\times G_{nl\vQ-\vq}^>(t,t')D_{\nu\vq}^<(t',t).
\label{eq:Fan}
\end{align}
The collision integral is a functional of the two-time Green's function $G^{>,<}(t,t')$.
Therefore, without further approximation, solving the density matrix dynamics would be equivalent to solve the full KBE.
To make progress, we employ the GKBA, which is a standard approximation of the two-time Green's function. 
Assuming $t>t'$, we have
\begin{align}
G^{<}(t,t') & =-G^{R}(t,t')\rho(t')\nonumber\\
G^{>}(t,t') & =G^{R}(t,t')\bar{\rho}(t')\nonumber\\
G^{<}(t',t) & =\rho(t')G^{A}(t',t)\nonumber\\
G^{>}(t',t) & =-\bar{\rho}(t')G^{A}(t',t)
\label{eq:GKBA}
\end{align}
, where $\bar{\rho}(t)=I+\rho(t)$ with the identity matrix $I$.
Substituting Green's function with the ansatz Eq.~\ref{eq:GKBA} in the collision integral, we obtain Eq.~\ref{eq:GKBA_scat} in the main text.
We proceed by using the Wigner-Weisskopf form for the propagators and further approximating the phonon propagator with their equilibrium form, ~\cite{QK2008}
\[
D_{\nu\mathbf{q}}^{>}(t,t')=-i\sum_{\alpha=\pm}N_{\nu\mathbf{q}}^{\alpha}e^{-i\alpha\omega_{\nu\mathbf{q}}(t-t')}
\]
and
\[
D_{\nu\mathbf{q}}^{<}(t,t')=-i\sum_{\alpha=\pm}N_{\nu\mathbf{q}}^{\alpha}e^{i\alpha\omega_{\nu\mathbf{q}}(t-t')}
\]
where $N_{\nu\vq}^{\pm}=n_{\nu\vq}+\frac{1}{2}\pm\frac{1}{2}$.
Equation~\ref{eq:GKBA_scat} can be reduced to a form analogous to the Lindblad framework by invoking the Markov approximation and extending the time-integration limit to $t \to \infty$~\cite{Marini_2013}. We emphasize that the basis dependence inherent in the Lindblad approach is a direct mathematical consequence of this infinite-time integration; the resulting collision operator becomes tied to the specific representation in which the energy-conserving delta function is enforced, thereby breaking the form-invariance of the dynamics.

\subsection{G1-G2 scheme}
\label{sec:g1g2}
The GKBA facilitates a numerically efficient scheme that reduces the computational complexity of evaluating Eq.~\ref{eq:GKBA_scat} from a scheme that scales as $O(t^2)$ down to $O(t)$~\cite{Bonitz2020,Perfetto2023}.
This is achieved by introducing a set of the auxiliary density matrices, defined as
\begin{widetext}
\begin{align*}
\mathcal{F}_{n\mathbf{Q-q}p\mathbf{Q}}^{\nu\pm<,>}(t) & =\mathcal{G}_{jl,\mathbf{Q}-\mathbf{q},\mathbf{q}}^{\nu*}\int_{0}^{t}dt'G_{ns\mathbf{Q}-\mathbf{q}}^{R}(t,t')\bar{\rho}_{sl\mathbf{Q}-\mathbf{q}}(t')D_{\nu\mathbf{q}}^{\pm<,>}(t',t)\rho_{jm\mathbf{Q}}(t')G_{mp\mathbf{Q}}^{A}(t',t),
\end{align*}
\end{widetext}
where $D^{>,<\pm}$ represent the $\alpha=\pm$ branch of the phonon propagator.
Within this framework, the collision integral Eq.~\ref{eq:GKBA_scat} can be recast in terms of these auxiliary matrices,
\begin{align*}
I_{ip\vQ}^{L<}(t)=&i\sum_{\nu\mathbf{q},\alpha=\pm}\mathcal{G}_{\nu in}(\vQ-\vq,\vq)\left[\mathcal{F}_{n\mathbf{Q-q}p\mathbf{Q}}^{\nu,\alpha,<}(t)\right.\\
&\left.-\mathcal{F}_{n\mathbf{Q-q}p\mathbf{Q}}^{\nu,\alpha,>}(t)\right]
\end{align*}
By taking the time derivative of the auxiliary density matrix, we obtain a closed equation of motion
\begin{align*}
i\frac{d}{dt}\mathcal{F}_{n\mathbf{Q-q}p\mathbf{Q}}^{\nu+<}(t) & =h_{nm}(t)\mathcal{F}_{m\mathbf{Q-q}p\mathbf{Q}}^{\nu+<}(t)-\mathcal{F}_{n\mathbf{Q-q}b\mathbf{Q}}^{\nu+<}(t)h_{bp}(t)\\
 & +\omega_{\nu\mathbf{q}}\mathcal{F}_{n\mathbf{Q-q}p\mathbf{Q}}^{\nu+<}(t)\\
 & +N_{\nu\mathbf{q}}^{+}\mathcal{G}_{\nu jl}(\mathbf{Q}-\mathbf{q},\mathbf{q})^{*}\bar{\rho}_{nl\mathbf{Q}-\mathbf{q}}(t)\rho_{jp\mathbf{Q}}(t).
\end{align*}
The equation of motion for the remaining three auxiliary density matrices are derived similarly.
This transformation converts the integro-differential equation of the density matrix into a coupled system of first-order ordinary differential equations, effectively 'mapping' the memory kernel onto the dynamics of the auxiliary variables.
In the numerical implementation, storage of the auxiliary density matrix becomes a bottleneck.
We take advantages of large scale parallelization to overcome this issue.

While this approach bypasses the unfavorable $O(t^2)$ scaling, the high dimensionality of the auxiliary matrices—which must be stored for each $\vQ$-point and exciton and phonon indices—presents a significant memory bottleneck. 
We address this through large-scale distributed-memory parallelization, enabling the simulation of full Brillouin zone dynamics.

\begin{figure*}[t]
     \centering
     \includegraphics[width=\textwidth]{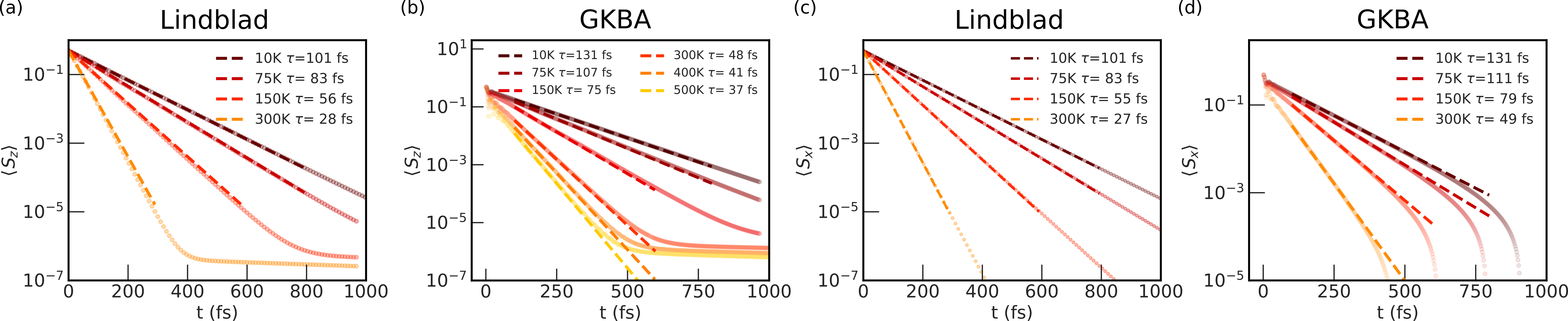}
\caption{Temperature-dependent valley dynamics. Computed valley depolarization (panels a, b) and decoherence (panels c, d) dynamics across a temperature range of $10\text{--}500\ \text{K}$, comparing the Lindblad and GKBA frameworks. Symbols represent the calculated data points, while dashed lines indicate the corresponding mono-exponential fits used to extract the characteristic decay times $\tau$. In the GKBA results, note the presence of non-Markovian transients during the initial $50\ \text{fs}$ that precede the onset of the exponential relaxation regime.}
\label{fig:detaileddynamics}
\end{figure*}

\section*{Appendix II: Temperature dependent valley dynamics}
\label{sec:appendix2}

Figure~\ref{fig:detaileddynamics} illustrates the temperature dependence of the valley dynamics calculated as calculated within the Lindblad and GKBA frameworks.
Both valley depolarization (Fig.~\ref{fig:detaileddynamics} (a)) and decoherence (Fig.~\ref{fig:detaileddynamics} (c)) dynamics exhibit mono-exponential decay in the Lindblad simulations across all temperatures.
In contrast, the GKBA results feature high-frequency oscillations during the initial 50 fs, which is subsequently transition into a mono-exponential decay regime.
The amplitude of these non-Markovian transients increases with temperature, suggesting that their magnitude scales with the total scattering rate-a consequence of the increased phonon population at elevated temperature. 

\section*{References}
\bibliography{ref}

\end{document}